# Bandwidth-Controlled Insulator-Metal Transition and Correlated Metallic State in 5*d* Transition Metal Oxides Sr$_{n+1}$Ir$_n$O$_{3n+1}$ (*n*=1, 2, and ∞)

S.J. Moon,[1] H. Jin,[2] K.W. Kim,[3] W.S. Choi,[1] Y.S. Lee,[4] J. Yu,[2] G. Cao,[5] A. Sumi,[6] H. Funakubo,[6] C. Bernhard,[3] and T.W. Noh[1,*]

[1]*ReCOE & FPRD, Department of Physics and Astronomy, Seoul National University, Seoul 151-747, Korea*

[2]*CSCMR & FPRD, Department of Physics and Astronomy, Seoul National University, Seoul 151-747, Korea*

[3]*Department of Physics and Fribourg Center for Nanomaterials, University of Fribourg, CH-1700 Fribourg, Switzerland*

[4]*Department of Physics, Soongsil University, Seoul 156-743, Korea*

[5]*Department of Physics and Astronomy, University of Kentucky, Lexington, Kentucky 40506, USA*

[6]*Deparment of Innovative and Engineered Materials, Interdisciplinary Graduate School of Science and Engineering, Tokyo Institute of Technology, Yokohama 226-8502, Japan*

We investigated the electronic structures of the 5*d* Ruddlesden-Popper series Sr$_{n+1}$Ir$_n$O$_{3n+1}$ (*n*=1, 2, and ∞) using optical spectroscopy and first-principles calculations. As 5*d* orbitals are spatially more extended than 3*d* or 4*d* orbitals, it has been widely accepted that correlation effects are minimal in 5*d* compounds. However, we observed a bandwidth-controlled transition from a Mott insulator to a metal as we increased *n*. In addition, the artificially synthesized perovskite SrIrO$_3$ showed a very large mass enhancement of about 6, indicating that it was in a correlated metallic state.

*E-mail address: twnoh@snu.ac.kr

Mott physics, driven by electron correlation, has been an essential foundation for understanding the emergent properties of strongly correlated electron systems, including 3*d* and 4*d* transition metal oxides (TMOs) [1]. According to the Hubbard model, such systems can be characterized by the on-site Coulomb interaction *U* and the bandwidth *W*. When *W*<<*U*, the system is in an insulating state, the so-called Mott insulator. When *W*≥*U*, an insulator-metal transition (IMT) occurs and the system becomes metallic. Close to the IMT, the conducting carriers have large effective mass due to electron correlation. This correlated metallic state (CMS) has unique physical properties, which cannot be explained by simple band theory. Therefore, the *W*-controlled IMT and associated CMS, especially in 3*d* and 4*d* TMOs, have been important subjects in the condensed matter physics [1-4].

The 5*d* orbitals are spatially more extended than the 3*d* and 4*d* orbitals, so their *W* (*U*) values should be larger (smaller) than those of 3*d* and 4*d* TMOs. And the electron correlation should play much less important role. Therefore, many 5*d* TMOs have metallic ground states that can be described by band theory [5, 6]. However, some 5*d* TMOs, such as $Sr_2IrO_4$, $Sr_3Ir_2O_7$, and $Ba_2NaOsO_6$, have insulating ground states [7-9]. There have been some recent reports that correlation effects could be important for the 5*d* insulating TMO [9, 10]. Then, we can raise a couple of important questions. First, is it possible that a *W*-controlled IMT occurs in 5*d* TMOs? Second, can the associated CMS be found in such a 5*d* TMO system?

Recently, using angle-resolved photoemission and x-ray absorption spectroscopy techniques, we found that the unusual insulating state of $Sr_2IrO_4$ could be explained by cooperative interaction between electron correlation and spin-orbit (SO) coupling [11]. The SO coupling constant of 5*d* TMOs is about 0.3-0.4 eV [6], which is much larger than that of 3*d* TMOs (~ 20 meV) [12]. As the SO coupling constant of 5*d* TMOs is comparable to their *W* and *U* values, it should play an important role in their electronic structures [13]. The ground state of

Sr$_2$IrO$_4$ should be occupied by five $t_{2g}$ electrons. The large SO coupling can split the degenerate $t_{2g}$ orbitals into $J_{eff,1/2}$ and $J_{eff,3/2}$ bands [14]. The $J_{eff,1/2}$ bands can be very narrow due to a decreased hopping integral caused by the isotropic orbital and mixed spin characters [11]. Although the $U$ values of 5$d$ TMOs are smaller than those of 3$d$ or 4$d$ TMOs, the very narrow feature of the $J_{eff,1/2}$ bands allows an insulating gap to open due to correlation, i.e., SO coupling-triggered Mott gap. As shown schematically in Fig. 1(a), the $J_{eff,1/2}$ bands split into a lower Hubbard band (LHB) and an upper Hubbard band (UHB), opening a Mott gap.

The weakly correlated narrow band system can be a good starting point from which to search for the $W$-controlled IMT and associated CMS in 5$d$ TMOs. The number of neighboring Ir atoms, $z$, in Sr$_2$IrO$_4$ is 4. As $W$ should be proportional to $z$, we could increase $W$ by increasing the $z$ value. Among the Ruddlesden-Popper series Sr$_{n+1}$Ir$_n$O$_{3n+1}$ compounds, the $z$ values for Sr$_3$Ir$_2$O$_7$ and perovskite SrIrO$_3$ are 5 and 6, respectively. In nature, however, bulk SrIrO$_3$ has a hexagonal crystal structure at room temperature and atmospheric pressure. It transforms to the perovskite structure only at a higher pressure and temperature [15]. Therefore it is essential to obtain a perovskite SrIrO$_3$ for the investigation.

In this Letter, we investigated the electronic structures of Ruddlesden-Popper series Sr$_{n+1}$Ir$_n$O$_{3n+1}$ ($n$=1, 2, and $\infty$) compounds using optical spectroscopy and first-principles calculations. We artificially fabricated the perovskite SrIrO$_3$ phase by growing an epitaxial thin film on a MgO substrate. We found that the $W$ of Ir 5$d$ bands increased with the $z$ value. As schematically shown in Figs. 1(b) and 1(c), $W$-controlled IMT can occur between Sr$_3$Ir$_2$O$_7$ and SrIrO$_3$. From optical conductivity spectrum $\sigma(\omega)$, we also found that conducting carriers in the artificially fabricated SrIrO$_3$ should have a large mass enhancement, indicating a CMS.

Single crystals of Sr$_2$IrO$_4$ and Sr$_3$Ir$_2$O$_7$ were grown using the flux technique. Transport and magnetic measurements showed that they had insulating ground states with weak ferromagnetism [7, 8]. To obtain a perovskite $SrIrO_3$ sample, we grew epitaxial $SrIrO_3$ thin film on cubic MgO (001) substrate. The cubic symmetry of the MgO substrate could ensure that the $SrIrO_3$ film grow in the perovskite phase [16]. X-ray diffraction pattern, shown in the inset of Fig. 2(c), showed that the $SrIrO_3$ film was grown epitaxially in the perovskite phase and that its *c* axis lattice constant was nearly the same as that of the high pressure and temperature bulk perovskite phase. The epitaxial stabilization of perovskite $SrIrO_3$ phase enabled us to investigate the *W*-controlled changes in the electronic structures of the iridates systematically.

We measured *ab* plane reflectance spectra of $Sr_2IrO_4$ and $Sr_3Ir_2O_7$ single crystals at room temperature. The corresponding $\sigma(\omega)$ were obtained using the Kramers-Kronig transformation. For the $SrIrO_3$ thin films, we measured reflectance and transmittance spectra in the energy region between 0.15 and 6 eV. By solving the Fresnel equations numerically [17], we obtained $\sigma(\omega)$ for $SrIrO_3$. Between 0.01 and 0.09 eV, we used far-infrared ellipsometry technique to obtain $\sigma(\omega)$ directly [18]. Due to the strong phonon absorption of the MgO substrate, we could not obtain $\sigma(\omega)$ between 0.09 and 0.15 eV.

Figure 2 shows that the $Sr_{n+1}Ir_nO_{3n+1}$ compounds experienced IMT when the *z* value, i.e., the *W* value, was increased. For the *z*=4 compound $Sr_2IrO_4$, $\sigma(\omega)$ showed a finite sized optical gap. For the *z*=5 compound $Sr_3Ir_2O_7$, the optical gap became almost zero. For the *z*=6 compound $SrIrO_3$, it finally disappeared, and a Drude-like response from conducting carriers appeared in the low frequency region. These spectral changes indicate that an IMT occurs between $Sr_3Ir_2O_7$ and $SrIrO_3$.

It should be noted that $\sigma(\omega)$ of $Sr_2IrO_4$ had unique spectral features [19], which are difficult to be found in most other Mott insulators. It has a double peak structure, marked as $\alpha$ and $\beta$ in Fig. 2(a). As shown in Fig. 1(a), the peak $\alpha$ can be assigned as an optical transition from the LHB to the UHB of the $J_{eff,1/2}$ states, while the peak $\beta$ can be assigned as that from the

$J_{eff,3/2}$ bands to the UHB. Note that the peak $\alpha$ is very narrow, whose width is 3-5 times smaller than those of the correlation-induced peaks in other 3$d$ and 4$d$ Mott insulators. For comparison, in Fig. 2(a), we plotted the reported $\sigma(\omega)$ of LaTiO$_3$ and Ca$_2$RuO$_4$ [20, 21], which are well-known Mott insulators. The sharpness of the peak $\alpha$ originated from the narrowness of the SO coupling triggered $J_{eff,1/2}$ Hubbard bands, which could facilitate $W$-controlled IMT with a small $U$ value of a 5$d$ TMO.

To gain insight into the electronic structure changes of Sr$_{n+1}$Ir$_n$O$_{3n+1}$, we performed local density approximation (LDA)+$U$ calculations including SO coupling [22]. We used a $U$ value of 2.0 eV, which produced electronic structures consistent with our experimental $\sigma(\omega)$. It is worthwhile to mention that most band structure calculations on 3$d$ Mott insulators usually took a $U$ value of 4-7 eV [12]. Figures 3(a), 3(b), and 3(c) show the band dispersions of Sr$_2$IrO$_4$, Sr$_3$Ir$_2$O$_7$, and SrIrO$_3$, respectively. In the energy region between -1.5 and 1.0 eV, the Ir 5$d$ states mainly contribute. The red and black lines represent the $J_{eff,1/2}$ bands and the $J_{eff,3/2}$ bands, respectively. For Sr$_3$Ir$_2$O$_7$ and SrIrO$_3$, the $J_{eff,1/2}$ bands split due to the increase in interlayer coupling. When $z$ increases, the neighboring Ir ions along the $c$ axis will have stronger hybridization of the $d$ bands through the apical O ions. The hybridization can split the bands into bonding and antibonding states, which resulted in an increase in $W$.

Figures 3(d), 3(e), and 3(f) show the total density of states (DOS) of Sr$_2$IrO$_4$, Sr$_3$Ir$_2$O$_7$, and SrIrO$_3$, respectively. The DOS between -0.5 and -1.5 eV had strong contribution from the $J_{eff,3/2}$ states. In Figs. 3(d) and 3(e), the DOS between 0 and 0.5 eV was from the UHB of the $J_{eff,1/2}$ states, while that between 0 and -0.5 eV came from the LHB of the $J_{eff,1/2}$ states. In Fig. 3(f), the DOS between -0.5 and 0.5 eV was from the $J_{eff,1/2}$ bands. For Sr$_2$IrO$_4$, the separation between the centers of the UHB and LHB ($J_{eff,3/2}$ bands) was approximately 0.5 eV (1.0 eV), consistent with the position of the peak $\alpha$ ($\beta$) in Fig. 2(a). As $z$ increased, the $J_{eff,1/2}$ and $J_{eff,3/2}$

bands clearly broadened. The $W$ values of the $J_{eff,1/2}$ bands in DOS were estimated to be about 0.48, 0.56, and 1.01 eV. As the $J_{eff,1/2}$ bands broadened, the Mott gap closed. These first-principles calculations demonstrate the nature of the $W$-controlled IMT in $Sr_{n+1}Ir_nO_{3n+1}$.

Experimental evidence for the systematic $W$ changes of the $J_{eff,1/2}$ bands was obtained from $\sigma(\omega)$. As shown in Fig. 2, the $\alpha$ and $\beta$ peaks broadened and decreased in energy with increase of $z$. For quantitative analysis, we fitted $\sigma(\omega)$ using the Lorentz oscillator model. For the insulators, we used three Lorentz oscillators, which corresponded to the peaks $\alpha$, $\beta$, and a charge transfer excitation from the O $2p$ bands to the Ir $5d$ bands. For $SrIrO_3$, we used the Drude model for metallic response and two Lorentz oscillators which corresponded to the peak $\beta$ and the charge transfer excitation. From this analysis, we obtained the peak positions and widths of the $\alpha$ and $\beta$ peaks, i.e., $\omega_\alpha$, $\omega_\beta$, $\gamma_\alpha$, and $\gamma_\beta$. According to the Fermi's golden rule, $\sigma(\omega)$ should be proportional to a matrix element and a joint density of states. Therefore, the width of an absorption peak should reflect the $W$ of the initial and final bands. As shown in Fig. 4(a), both the $\gamma_\alpha$ and $\gamma_\beta$ values increased as the $z$ value became larger. This confirmed that the IMT in $Sr_{n+1}Ir_nO_{3n+1}$ should be $W$-controlled.

The Lorentz oscillator model analysis provided further information on the electronic structure changes. As shown in Fig. 4(a), both the $\omega_\alpha$ and $\omega_\beta$ values decreased systematically as $z$ was increased. For the $\omega_\alpha$ change, the electronic structure changes for both LHB and UHB are important. However, the $\omega_\beta$ change only involves the changes in the UHB. In Fig. 4(a), the change in $\omega_\alpha$ between $Sr_2IrO_4$ and $Sr_3Ir_2O_7$ was approximately 0.25±0.05 eV. On the other hand, the change in $\omega_\beta$ was about 0.13±0.05 eV, which is approximately half of the change in $\omega_\alpha$. This suggested that the electronic structure changes should occur symmetrically around the Fermi level for both Hubbard bands, as shown in Fig. 1.

To obtain a firm evidence for Mott instability in $5d$ $Sr_{n+1}Ir_nO_{3n+1}$, we performed extended Drude model analysis on metallic SrIrO$_3$ [23, 24]. From the optical spectra, we obtained the mass enhancement of conducting carriers $\lambda(\omega)$, which corresponded to the effective mass normalized by the band mass. Substantial enhancement of $\lambda(\omega)$ near the IMT is a hallmark of Mott instability and the corresponding compound should be in CMS. As shown in Fig. 4(b), the $\lambda(\omega)$ of SrIrO$_3$ reached about 6 at the lowest energy. We are not aware of any report on $\lambda(\omega)$ for 5$d$ TMOs, so we included the room temperature $\lambda(\omega)$ of 4$d$ correlated metals SrRuO$_3$ and Sr$_2$RuO$_4$ in Fig. 4(b) for comparison. The $\lambda(\omega)$ of 3$d$ correlated metal (Ca, Sr)VO$_3$ was reported to be approximately 3-4 [25]. Note that the $\lambda(\omega)$ of SrIrO$_3$ is comparable to or larger than those of the 3$d$ and 4$d$ TMOs. It is quite interesting that such a large value of $\lambda(\omega)$ was found for the 5$d$ TMO compound. This large value implies that SrIrO$_3$ is a correlated metal close to the Mott transition.

Our findings, the $W$-controlled IMT and the large effective mass of the resulting metallic compound in 5$d$ system, challenge the conventional expectation that the electron correlation is insignificant in 5$d$ system. These unique findings originate from the large SO coupling of 5$d$ transition metal ions. Recent theoretical results showed that the SO coupling in 4$d$ counterpart compound could just modify the Fermi surface within the metallic phase [26, 27]. On the other hand, the large SO coupling in 5$d$ system could drastically enhance the effect of the electron correlation and this cooperative interaction drives some 5$d$ system close to the Mott instability.

In summary, we investigated the electronic structures of a weakly correlated narrow band system, 5$d$ Sr$_{n+1}$Ir$_n$O$_{3n+1}$, using optical spectroscopy and first-principles calculations. We observed a bandwidth-controlled insulator-metal transition and the associated correlated metallic state with a large effective mass. Our results clearly demonstrate that the electron correlation could play an important role even in 5$d$ systems.

This work was financially supported by the Creative Research Initiative Program (Functionally Integrated Oxide Heterostructure) of MOST/KOSEF and Schweizer Nationalfonds with grant 200020-119784. YSL was supported by the Soongsil University Research Fund. The experiments at PLS were supported in part by MOST and POSTECH.

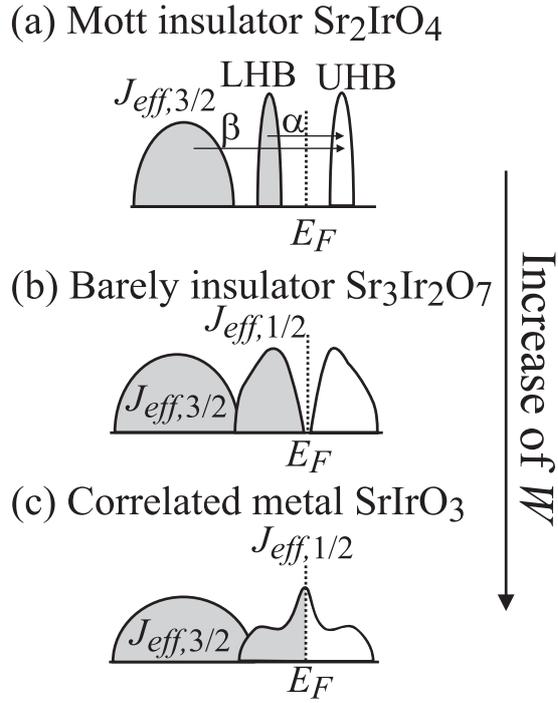

Fig. 1. Schematic band diagrams of 5$d$ Sr$_{n+1}$Ir$_n$O$_{3n+1}$ compounds, which are well described by the effective total angular moment $J_{eff}$ states due to strong spin-orbit coupling: (a) Mott insulator Sr$_2$IrO$_4$, (b) barely insulator Sr$_3$Ir$_2$O$_7$, and (c) correlated metal SrIrO$_3$. $E_F$ represents the Fermi level and the arrow indicates the direction for the bandwidth $W$ increase.

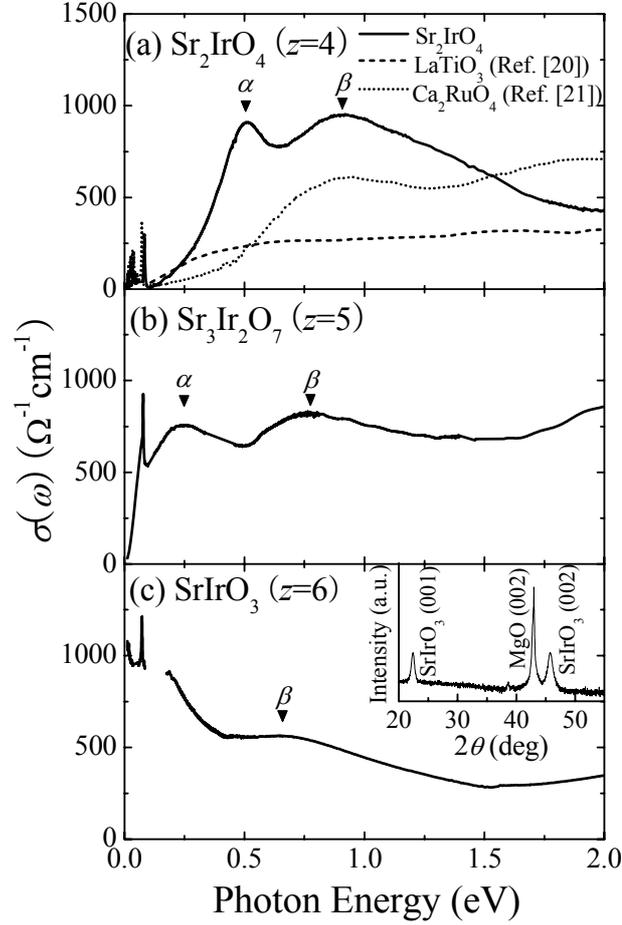

Fig. 2. Optical conductivity spectra $\sigma(\omega)$ of (a) $Sr_2IrO_4$, (b) $Sr_3Ir_2O_7$, and (c) $SrIrO_3$. In (a), $\sigma(\omega)$ of other Mott insulators, such as $3d$ $LaTiO_3$ [Ref. 20] and $4d$ $Ca_2RuO_4$ [Ref. 21], are shown for comparison. The peak $\alpha$ corresponds to the optical transition from the LHB to the UHB of the $J_{eff,1/2}$ states. The peak $\beta$ corresponds to the transition from the $J_{eff,3/2}$ bands to the UHB. The inset of (c) shows the x-ray diffraction pattern of the perovskite $SrIrO_3$ film, which was artificially synthesized on a MgO(001) substrate.

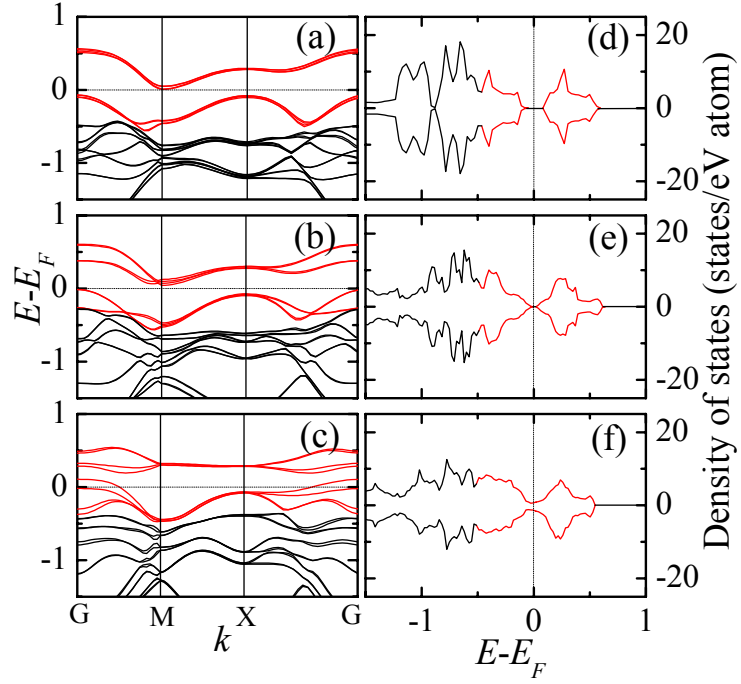

Fig. 3 (color online). Results from the LDA+$U$ calculations including spin-orbit coupling: Band structures of (a) $Sr_2IrO_4$, (b) $Sr_3Ir_2O_7$, and (c) $SrIrO_3$, and total density of states of (d) $Sr_2IrO_4$, (e) $Sr_3Ir_2O_7$, and (f) $SrIrO_3$. The positive and negative densities of states values represent spin-up and spin-down bands, respectively. The red and black lines represent the $J_{eff,1/2}$ and the $J_{eff,3/2}$ bands, respectively.

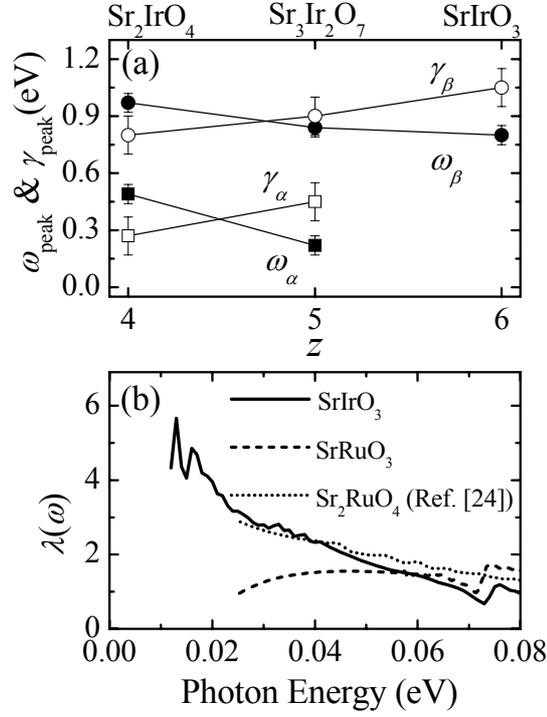

Fig. 4. (a) Results from Lorentz oscillator model analysis. The solid symbols show the positions of the $\alpha$ and $\beta$ peaks, i.e., $\omega_\alpha$ and $\omega_\beta$. The open symbols indicate the widths of the $\alpha$ and $\beta$ peaks, i.e., $\gamma_\alpha$ and $\gamma_\beta$. (b) Frequency-dependent mass enhancement $\lambda(\omega)$ of correlated metallic $SrIrO_3$. The $\lambda(\omega)$ data of $SrRuO_3$ (our data) and $Sr_2RuO_4$ (Ref. 24) are included for comparison.